\documentclass{article}
\PassOptionsToPackage{numbers}{natbib}
\usepackage[preprint]{neurips_2022}

\usepackage[utf8]{inputenc} 
\usepackage[T1]{fontenc}    
\usepackage{hyperref}       
\usepackage{url}            
\usepackage{booktabs}       
\usepackage{amsfonts}       
\usepackage{nicefrac}       
\usepackage{microtype}      
\usepackage{xcolor}         
\usepackage{graphicx}
\usepackage{siunitx}

\graphicspath{{./images/}}

\newcommand{\cond}[2]{#1 \,|\, #2}
\newcommand{\vvar}[1]{{\bf #1}}
\newcommand{\pgen}{p_{\theta}}
\newcommand{\qrec}{q_{\Phi}}

\title{Top-down inference in an early visual cortex inspired hierarchical Variational Autoencoder}

\author{%
  Ferenc Csikor$^*$\\
  Department of Computational Sciences\\
  Wigner Research Centre for Physics\\
  Budapest, Hungary \\
  \texttt{ferenc.csikor@gmail.com} \\
\And
  Bal\'azs Mesz\'ena\thanks{equal contributions} \\
  Department of Computational Sciences\\
  Wigner Research Centre for Physics\\
  Budapest, Hungary \\
  \texttt{meszenab@gmail.com} \\
\And
  Bence Szab\'o \\
  Department of Computational Sciences\\
  Wigner Research Centre for Physics\\
  Budapest, Hungary \\
  \texttt{bszabo96@gmail.com} \\
\And
  Gerg\H{o} Orb\'an \\
  Department of Computational Sciences\\
  Wigner Research Centre for Physics\\
  Budapest, Hungary \\
  \texttt{orgergo@gmail.com} \\
}

\begin{document}

\maketitle

\begin{abstract}
  Interpreting computations in the visual cortex as learning and inference in a generative model of the environment has received wide support both in neuroscience and cognitive science. However, hierarchical computations, a hallmark of visual cortical processing, has remained impervious for generative models because of a lack of adequate tools to address it. Here we capitalize on advances in Variational Autoencoders (VAEs) to investigate the early visual cortex with sparse coding hierarchical VAEs trained on natural images. We design alternative architectures that vary both in terms of the generative and the recognition components of the two latent-layer VAE. We show that representations similar to the one found in the primary and secondary visual cortices naturally emerge under mild inductive biases. Importantly, a nonlinear representation for texture-like patterns is a stable property of the high-level latent space resistant to the specific architecture of the VAE, reminiscent of the secondary visual cortex. We show that a neuroscience-inspired choice of the recognition model, which features a top-down processing component is critical for two signatures of computations with generative models: learning higher order moments of the posterior beyond the mean and image inpainting. Patterns in higher order response statistics provide inspirations for neuroscience to interpret response correlations and for machine learning to evaluate the learned representations through more detailed characterization of the posterior.
\end{abstract}

\section{Introduction}
\label{sec:intro}

Interpreting visual perception as inference in an unsupervised generative model is a key concept in neuroscience \cite{yuille2006vision,fiser2010statistically}. Specifically, using \emph{generative} models of natural images to understand the response statistics of neurons in the visual cortex of mammals has proven a lucrative approach. Linear (or close to linear) models have been instructive to account for a wide spectrum of the response characteristics of the neurons found in the primary visual cortex (V1), such as receptive fields structure \cite{olshausen1996emergence,karklin2009emergence}, extra-classical receptive field properties \cite{schwartz2001natural}, or response variability \cite{hoyer2002interpreting,orban2016neural,echeveste2020cortical,festa2021neuronal}. However, going beyond linear models has been largely hampered by limits on the capabilities of machine learning tools to perform learning and inference in nonlinear generative models. Recently, nonlinear generative models have attracted considerable attention \cite{kingma2013auto,rezende2014stochastic,goodfellow2014generative,kingma2019introduction}. Variational Autoencoders (VAEs) are a flexible class of deep generative models which use deep neural networks to parametrize highly nonlinear computations necessary for discovering  nonlinear and potentially disentangled latent features as well as delicate invariances characteristic of higher level visual areas \cite{dicarlo2007untangling,zoccolan2007trade,burgess2018understanding,higgins2021unsupervised,ziemba2016selectivity}. Surprisingly, the application of VAEs to neural computations has been limited \cite{higgins2021unsupervised,barello2018sparse}.

The visual cortex of primates and other mammals is characterized by a hierarchy of processing stages. Consequently, establishing links between deep generative models and visual cortical computations requires hierarchical versions of VAEs (hVAEs). Recently, considerable advances were made in constructing effective hVAE architectures \cite{maaloe2019biva,vahdat2020nvae,zhao2017infovae,child2020very,sonderby2016ladder, hazami2022efficient}. Despite the recent successes in generating high quality images with hVAEs, the representations learned by hVAEs are less studied \cite{chen2021boxhead}. From a machine learning perspective, constructing hVAEs has multiple challenges, which stem from the unidentifiability of the latent representation, instabilities of training, sensitivity to the complexity of the generative model, and collapse of latent representations \cite{sonderby2016ladder,dai2020usual, khemakhem2020variational}. As a consequence, alternative hVAE architectures are distinguished by the way they can cope with the above challenges. Besides the efficiency of training, more principled considerations might guide us when constructing hVAEs. The motivation for learning hVAEs is twofold. First, a hierarchical representation offered by it has the potential to provide increasingly more complex features that the VAE can infer from observations. A cascade of such features enables rich and flexible inferences and can therefore support a larger variety of tasks. Second, an inherent constraint of variational inference is that the posterior has a simple form, and it is very limited in its capacity to express dependencies between features. A single latent-layer VAE is thus constrained by the expressivity of the variational posterior. An additional layer is a principled solution to introduce rich posterior structure through constraints imposed by high-level latent variables. In this paper we investigate how machine learning principles can be aligned with neuroscience motivations to construct hVAEs.

We harness motivations coming from neuroscience to investigate the properties of representations and inference in a hVAE, which we list below. We are using a two latent-layer VAE that can be related to the primary and secondary visual cortices (V1 and V2). VAEs use a pair of models, the generative and the recognition  model, the first to learn the likelihood and the second for computing amortized variational posteriors for inference. In a hVAE, architectural choices concern both the generative and the recognition components: skip connections in the generative model that link higher layers of latent variables to the observed variables can stabilize learning and establish a richer dependency structure. However, it is less well motivated from an interpretable representation learning and neuroscience perspective than a Markovian generative model where dependencies are defined only between immediately neighboring layers of variables. Skip connections in the recognition model establish a direct link from the observed layer to the topmost latent layer, and lead to an alternative formulation of the training objective as a chain-like recognition model where information is propagating in a feed-forward manner from layer to layer. The recognition model that features a skip connection is also sporting a top-down component, which is well motivated by the anatomy of the visual cortex. Inspired by the success of linear generative models in V1 \cite{olshausen1996emergence,bell1997independent,schwartz2001natural}, we constrain the first layer of the generative model to be linear. This way, we both have better intuitions about the emerging representations, and it makes it possible to investigate the properties of inference in more detail. An additional neuroscience motivation concerns the choice of the prior: emerging representations are explored both under the standard normal prior and also under a sparser, Laplacian prior. The last motivation concerns the construction of data sets used both for training and testing: to match the conditions a biological system is adapted to, we use natural images for training and also use synthetic texture images for testing, since neurons in V2  display texture sensitivity \cite{freeman2013functional,ziemba2016selectivity, freeman2011metamers}.

In this paper we first develop a biologically inspired form of hVAE, TDVAE. Next, we introduce a variety of architectures that represent alternative forms for the generative and recognition models. We compare these architectures based on the representations they learn. We point out that in contrast to the general approach to compare representations based on the \emph{maximum a posteriori} inferences, probabilistic inference implies patterns in higher order moments as well, including posterior correlations. Comparison of architectures based on learned correlation yields insights into the higher efficiency inference of some of the architectures. We find that hVAE architectures are capable of reproducing some key properties of representations emerging in V1 and V2 of the visual cortex. Further, we find that architectures that feature top-down influences in their recognition model feature a richer representation, such that specific information that is not present in mean activations becomes linearly decodable from posterior correlations. Finally, we have found superior image inpainting properties compared to a single layer VAE via the reconstruction of masked natural images without explicit training for this task.

\section{Methods}
\label{sec_meth}

\textbf{Generative model}\quad
We investigate the question of learning the distribution of natural image data, $p({\bf x})$, via a hierarchical latent variable generative model, where we learn the joint distribution of observations and latent variables, $p_{\theta}({\bf x},  {\bf z}_1, {\bf z}_2, \dots {\bf z}_K)$. From an unsupervised learning perspective, a critical role of higher-order latent variables is to shape the posterior distribution of lower-level latent variables (similarly, a contextual prior can be established by high-level variables for low-level latent variables).

The focus of the paper will be on two-layer hierarchical models therefore we will omit latents higher than ${\bf z}_2$ from the discussion. The general form of such hierarchical model is
\begin{equation}
    p_{\theta}({\bf x}, {\bf z}_1, {\bf z}_2) = p_{\theta}({\bf x}\,|\, {\bf z}_1, {\bf z}_2) \cdot p_{\theta}({\bf z}_1\,|\, {\bf z}_2) \cdot p_{\theta}({\bf z}_2).
\end{equation}

We will especially be interested in generative models with a Markovian structure when $p_{\theta}({\bf x}\,|\, {\bf z}_1, {\bf z}_2) = p_{\theta}({\bf x}\,|\, {\bf z}_1)$. We will see that the representation learned by such a Markovian generative model is more interpretable as low level and higher level (more semantic) latent variables decouple.

Learning and inference in a nonlinear generative model are in general highly challenging. A framework that has been successful in addressing these challenges is the Variational Autoencoder (VAE) \cite{kingma2013auto, rezende2014stochastic}. In VAEs the generative model is supplemented by a recognition model, which establishes a variational approximation of the posterior distribution. In a two latent-layer setting the variational posterior  $\qrec(\cond{\vvar{z_1}, \vvar{z_2}}{\vvar{x}})$, approximates the true posterior $\pgen(\cond{\vvar{z_1}, \vvar{z_2}}{\vvar{x}})$.

In general, the ELBO (which is the difference between the log data likelihood and the difference between the true and the variational posterior) can be written for a two-layer generative model as
\begin{equation}
    {\rm ELBO}(\vvar{x}, \theta,\Phi) = {\rm E}_{\qrec(\cond{\vvar{z_1}, \vvar{z_2}}{\vvar{x}})}  [\pgen(\cond{\vvar{x}}{\vvar{z_1}, \vvar{z_2}})]  - {\rm KL} [\qrec(\cond{\vvar{z_1}, \vvar{z_2} }{\vvar{x}})\,||\, \pgen(\vvar{z_1}, \vvar{z_2})].
\end{equation}

\textbf{Factorization of the variational posterior}\quad
When dealing with a variational approximation of a hierarchical generative model, there are multiple choices to consider for the form of the recognition model. In practice we wish to factorize $\qrec(\cond{\vvar{z_1}, \vvar{z_2}}{\vvar{x}})$. An inspiration of how to do this can be obtained from the factorization properties of the true posterior. For the non-Markovian generative model one can factorize the posterior in a bottom-up fashion:
\begin{equation}
p(\cond{\vvar{z_1}, \vvar{z_2}}{\vvar{x}}) = p(\cond{\vvar{z_2}}{\vvar{x}, \vvar{z_1}}) \cdot
p(\cond{\vvar{z_1}}{\vvar{x}}). \label{eq:bufactor}
\end{equation}
For the Markovian case this simplifies to a chain as $p(\cond{\vvar{z_2}}{\vvar{x}, \vvar{z_1}})$ is independent of $x$.
For both the Markovian and non-Markovian case the posterior can be factorized in a top-down manner:
\begin{equation}
p(\cond{\vvar{z_1}, \vvar{z_2}}{\vvar{x}}) = p(\cond{\vvar{z_1}}{\vvar{x}, \vvar{z_2}}) \cdot
p(\cond{\vvar{z_2}}{\vvar{x}}). \label{eq:tdfactor}
\end{equation}
It is a reasonable choice to have a factorization of the variational posterior which preserves either of the above factorizations.
In the top-down case we will have a simple functional form (for example diagonal normal or Laplace distribution) for $\qrec(\cond{\vvar{z_2}}{\vvar{x}})$ and $\qrec(\cond{\vvar{z_1}}{\vvar{x}, \vvar{z_2}})$ and optimize for the TD objective function (for the Markovian generative model):
\begin{eqnarray}
    {\rm \mathcal{F}_{TD}}(\vvar{x}, \theta,\Phi) &=& {\rm E}_{\qrec(\cond{\vvar{x}}{\vvar{z}_2})\qrec(\cond{\vvar{z}_1}{\vvar{x}, \vvar{z}_2})}  [\pgen(\cond{\vvar{x}}{\vvar{z}_1})]  - \nonumber \\
    &-& \beta_1 \cdot {\rm E}_{\qrec(\cond{\vvar{z}_2}{\vvar{x}})} [{\rm KL} [\qrec(\cond{\vvar{z}_1}{\vvar{x}, \vvar{z}_2})\,||\, \pgen(\cond{\vvar{z}_1}{\vvar{z}_2})]] - \nonumber \\
    &-& \beta_2 \cdot {\rm KL} [\qrec(\cond{\vvar{z}_2}{\vvar{x}})\,||\, \pgen(\vvar{z}_2)].
    \label{eq:helbo}
\end{eqnarray}
For the non-Markovian case, the only difference is that in the reconstruction term $\pgen(\cond{\vvar{x}}{\vvar{z}_1})$ needs to be replaced by $\pgen(\cond{\vvar{x}}{\vvar{z}_1, \vvar{z}_2})$. Note that the KL term is a sum of layer wise KL terms. Recent deep hVAEs use this latent posterior structure (with a non-Markovian generative model) \cite{vahdat2020nvae,child2020very}.

We allowed to scale the KL terms with parameters $\beta_1$ and $\beta_2$. When $\beta_1=\beta_2=1$ the objective function is identical to the ELBO. On one hand it can help the training process by allowing these parameters to slowly increase from a small value to 1 ($\beta$ annealing). From a representation learning point of view $\beta$s can shape the latent representation by manipulating the mutual information between the observed and the latent variables \cite{burgess2018understanding, alemi2016deep}.

If we choose bottom-up factorization we constrain $\qrec(\cond{\vvar{z_1}}{\vvar{x}})$ to be of a simple form and the ELBO becomes for the Markovian case (since we only consider this type of generative model with the bottom-up recognition model):
\begin{eqnarray}
    {\rm \mathcal{F}_{BU}}(\vvar{x}, \theta,\Phi) &=& {\rm E}_{\qrec(\cond{\vvar{z}_1}{\vvar{x}})}  [\pgen(\cond{\vvar{x}}{\vvar{z}_1}) + \beta_1 \cdot \pgen(\cond{\vvar{z}_1}{\vvar{z}_2})]  - \beta_1 \cdot {\rm H}[\qrec(\cond{\vvar{z}_1}{\vvar{x}})] \nonumber \\
    &-& \beta_2 \cdot {\rm E}_{\qrec(\cond{\vvar{z}_1}{\vvar{x}})}{\rm KL} [\qrec(\cond{\vvar{z}_2}{\vvar{z}_1})\,||\, \pgen(\vvar{z}_2)],
    \label{eq:helbo2}
\end{eqnarray}
 where {\rm H} denotes the entropy of the distribution.

 Note that it is not compulsory to choose a factorization of the variational posterior which preserves Eq.~(\ref{eq:bufactor}) or (\ref{eq:tdfactor}). It would just mean that we can not saturate the ELBO even if we have a very expressive distribution for the single layer variational posteriors. For example in \cite{sonderby2016ladder} the authors choose to work with $\qrec(\cond{\vvar{z_1}, \vvar{z_2}}{\vvar{x}}) = \qrec(\cond{\vvar{z_1}}{\vvar{x}}) \cdot \qrec(\cond{\vvar{z_2}}{\vvar{x}})$. For them, the hierarchical nature of the representation comes from choosing a more compressing network for $\qrec(\cond{\vvar{z_2}}{\vvar{x}})$ compared to $\qrec(\cond{\vvar{z_1}}{\vvar{x}})$.

 According to the VAE framework, individual mappings between variables (the components of the generative and recognition models) are parameterised through neural networks. Non-linearities are implemented by multi-layer networks that consist of elementary units with the softplus activation function. We used fully connected neural networks instead of CNNs to avoid indirect effects of CNNs on the emerging representations.

 For the higher layer, we choose the standard normal distribution as the form of prior and posterior. For a variable modelling V1 activity ($\vvar{z}_1$ in our case) a well-known inductive bias is the sparseness of the prior \cite{olshausen1996emergence}. We will compare the two choices in the emerging $\vvar{z}_1$ representation but will use mostly the sparse version in the rest of the paper. For computational simplicity we decided to implement sparseness by jointly changing the distributions in the generative and the recognition models, the latter corresponding to an altered class of variational posterior distribution. We chose the Laplace distribution as a sparse replacement of the normal, which resulted in tractable KL-divergences.

 To impose a simple inductive bias on the lower layer of the generative model, we restricted it to a linear mapping: $\pgen(\cond{\vvar{x}}{\vvar{z}_1}) = \mathcal{N}(\vvar{x};\vvar{A}\vvar{z}_1, \vvar{I})$. We used an overcomplete representation to ensure that the linear model is not clipping information when performing inference, with a $\vvar{z_1}$ dimension 1.125 times the dimension of $\vvar{x}$.

\begin{figure}
  \centering
  \includegraphics[width=11.7cm, trim={0 7cm 0 2.5cm}, clip]{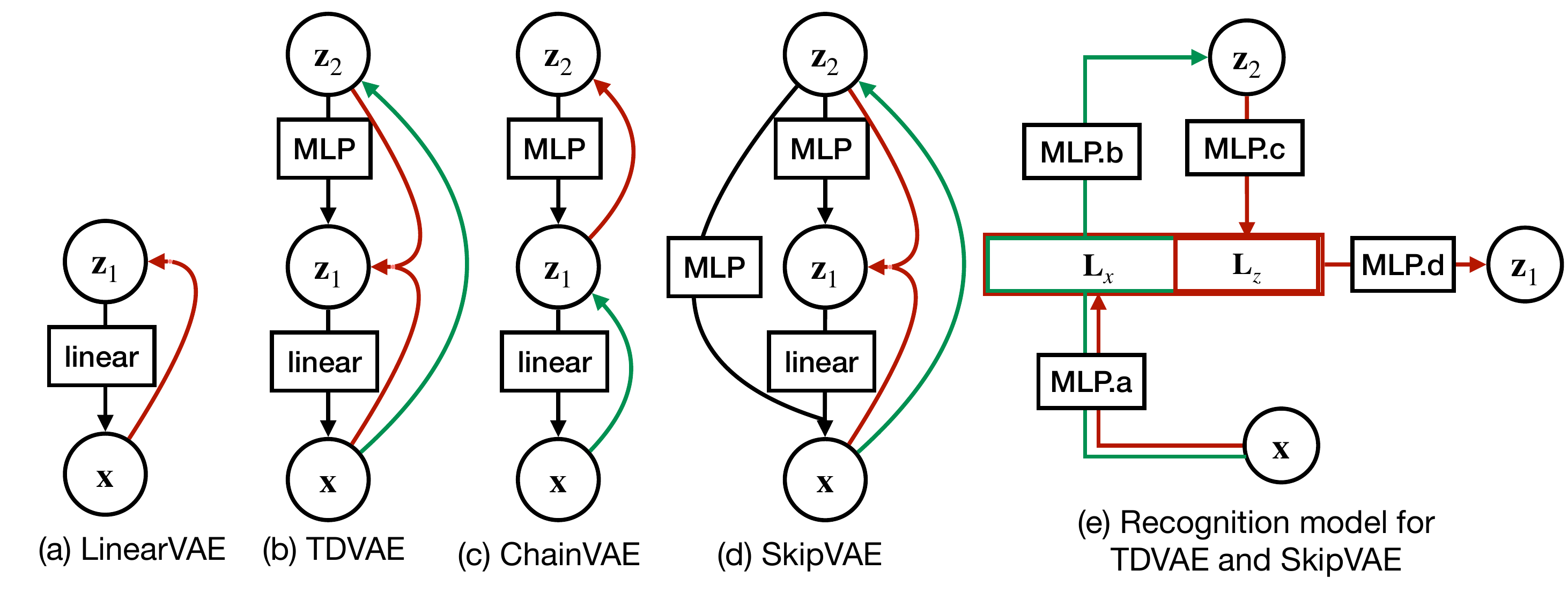}
  \caption{(a)-(d) The generative models used in the paper and schematic representation of the recognition models. (e) Detailed cartoon of the recognition model for TDVAE and SkipVAE. The green/red arrows highlight the computational flow for calculating the parameters of the distributions $\qrec(\cond{\vvar{z_2}}{\vvar{x}})$ / $\qrec(\cond{\vvar{z_1}}{\vvar{x}, \vvar{z_2}})$. The weights of MLP.a which produces the intermediate layer $L_x$ is shared between the two computations. The bottom-up and top-down components producing the $L_x$ and $L_z$ intermediate layers are concatenated to yield the input to MLP.d}
  \label{fig:gen_rec}
\end{figure}

\textbf{Architectures}\quad
In Fig. \ref{fig:gen_rec}(a)-(d) we summarized the high level architectures that are appearing in this paper. For easier reference we use the following terminology throughout the paper: a) LinearVAE is a single-layer VAE with linear generative model (this setup is identical to the one in \cite{barello2018sparse}); b) TDVAE is a two-layer VAE with Markovian generative model and top-down factorization of the variational posterior; c) ChainVAE is a two-layer VAE with Markovian generative model and bottom-up factorization of the posterior; and d) SkipVAE is a two-layer VAE with non-Markovian generative model and a top-down factorization of the posterior.

 The computational graph of the recognition model for the TDVAE and SkipVAE are non-trivial and contains further inductive biases (Fig. \ref{fig:gen_rec}(e)). We can see that there is a direct connection from $\vvar{x}$ to the stochastic variable $\vvar{z_2}$ in the sense that the latter does not depend on $\vvar{z_1}$. We emphasize however that this property is independent of the choice of the generative model and this form of recognition model is compatible with a Markovian generative model. More details about the recogntion model architecture can be found in the Supplementary Material.

\textbf{Stimuli}\quad
Training was performed on $20 \times 20$ and $40 \times 40$ pixel whitened natural image patches. We sampled \num{6.4e5} training and \num{6.4e4} test images from the van Hateren database \cite{van1998independent}, matched their grand total intensity histograms to the unit normal distribution, and applied the whitening procedure described in \cite{barello2018sparse}. This whitening procedure discards $100 (1 - \pi / 4) \%$ of the high frequency PCA components, keeping 314 and 1256 data dimensions in $20 \times 20$ and $40 \times 40$ image patches, respectively (see Fig.~\ref{fig:representation}a for examples).

We also used synthetic textures for testing purposes. Textures have a particular appeal since these are characterized by nonlinear sufficient statistics  \cite{portilla2000parametric, ustyuzhaninov2016does}, and biological systems show high sensitivity to this statistics \cite{freeman2013functional, ziemba2016selectivity}. We used five texture families that we selected based on the linear representation they gave rise to (we sought localized oriented filters).  Synthetic data sets were generated by an optimization algorithm developed by Portilla and Simoncelli\cite{portilla2000parametric} and whitened with the same procedure as the natural data sets (see Fig.~\ref{fig:representation}b for examples and Supplementary Material for more details). Seed images of the five texture families were downloaded from \url{https://www.textures.com/} under a license that granted free usage for noncommercial purposes, as well as from \cite{brodatz-1966-textures}.

\section{Related work}
\label{sec:relatedwork}

\textbf{Hierarchical representations and image completion with VAEs}\quad
Recently there have been significant progress in scaling up non-Markovian hVAEs \cite{vahdat2020nvae,child2020very,hazami2022efficient} but their learnt representations were much less studied. In \cite{chen2021boxhead} the authors have introduced a new dataset where the ground truth generative factors are known and have a hierarchical structure (by construction). Then, they studied the representation learned by a hVAE trained on this dataset. In this paper because of the vision/neuroscience motivation we have used natural images which are not synthetic but some properties are known. We have compared different representations and their interpretability emerging from various inductive biases.
In the paper \cite{harvey2021image} the authors have been able to achieve state of the art image completion results via deep generative models. They trained a model which can be described as a multimodal hVAE where the original and corrupted images are both shown at training time. We have also studied image inpainting on masked natural images and texture patches with our model. However, we did not train a dedicated model for the inpainting task (where the corrupted images are shown), rather studied reconstruction by our original model when presented with the masked images.

\textbf{Generative models and neural representations}\quad
Using generative models to understand biological vision has a strong tradition. In particular, sparse generative models have been successful in accounting for V1 responses \cite{olshausen1996emergence,bell1997independent,schwartz2001natural, hyvarinen2000emergence}. Despite arguments against shaping VAE representations through changing the prior \cite{miao2021incorporating}, sparse prior was shown to directly contribute to meaningful representations in a linear VAE \cite{barello2018sparse} but it it remains unclear if similar contribution  is achieved in hVAEs. Another VAE used strong nonlinearities to understand visual processing in a higher level of hierarchy \cite{higgins2021unsupervised}. This model was not hierarchical and training was more tailor-made by training on face images. A line of research that heavily relies on generative models are predictive coding models. Albeit these models explicitly implement generative models, their training objective is different from VAEs \cite{marino2022predictive}. In these models, instead of learning higher-order dependencies between low-level features, higher levels of hierarchy attempt to reconstruct activations at lower layers. These models have provided important insights into how learning dynamical representations contribute to neural phenomena \cite{lotter2020neural} and how contour integration can occur \cite{boutin2021sparse} but these do not touch upon learning expressive posteriors. Other studies that investigate higher order visual cortical representations \cite{hosoya2015hierarchical} also used a hierarchical model but without the probabilistic aspect that is essential in VAEs.

\section{Experiments}
\label{sec:experiments}

\begin{table}
  \caption{Models presented in this paper.}
  \label{tab:models}
  \centering
  \begin{tabular}{llrlrrrr}
    \toprule
    Name & Architecture & Patch size & $\vvar{z_1}$ distr.'s & $\textrm{dim}(\vvar{z_1})$ & $\textrm{dim}(\vvar{z_2})$ & $\beta_1$ & $\beta_2$ \\
    \midrule
    LinearVAE40   & LinearVAE & $40 \times 40$ & Laplace & 1800 & N/A   & 1.00 & N/A  \\
    TDVAE40       & TDVAE     & $40 \times 40$ & Laplace & 1800 & 250   & 1.00 & 1.00 \\
    TDVAE40n      & TDVAE     & $40 \times 40$ & normal  & 1800 & 250   & 1.00 & 1.00 \\
    TDVAE40125    & TDVAE     & $40 \times 40$ & Laplace & 1800 & 250   & 1.00 & 1.25 \\
    SkipVAE40     & SkipVAE   & $40 \times 40$ & Laplace & 1800 & 250   & 1.00 & 1.00 \\
    TDVAE20       & TDVAE     & $20 \times 20$ & Laplace & 450 & 5--140 & 1.00 & 1.00 \\
    SkipVAE20     & SkipVAE   & $20 \times 20$ & Laplace & 450 & 5--140 & 1.00 & 1.00 \\
    ChainVAE20    & ChainVAE  & $20 \times 20$ & Laplace & 450 & 70     & 1.00 & 1.00 \\
    \bottomrule
  \end{tabular}
\end{table}

We trained several model instances representing the model architectures in Section~\ref{sec_meth} on whitened natural image patches. The main hyperparameters of a selected subset of experiments is shown in~Table~\ref{tab:models}. Training the models took 727~hours on an internal computer supporting an Nvidia GeForce RTX 3080 Ti GPU. For further details and code please consult the Supplementary Material.

\subsection{Low-level representation}
\label{sec:low_level_repr}

\begin{figure}
  \centering
  \includegraphics[width=0.99\textwidth]{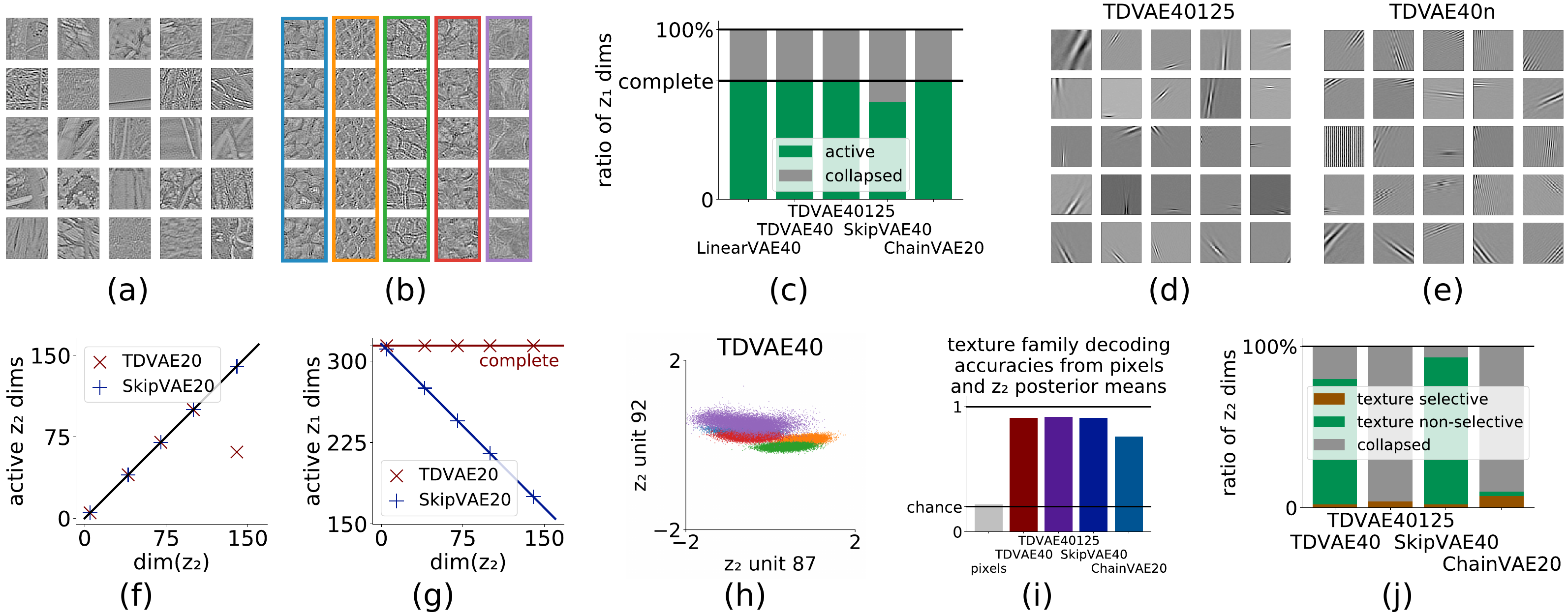}
  \caption{Characteristics of low- and high-level representations in natural image-trained hVAEs. (a)~Example $40 \times 40$ whitened natural image patches, with matching grey scale. (b)~Similarly, texture patches, \emph{colors} correspond to texture families. (c)~SkipVAE models learn undercomplete $\vvar{z_1}$ bases, all others learn complete ones. (d)~Gabor-like $\vvar{z_1}$ projective fields with Laplace generative and recognition distributions in $\vvar{z_1}$. Per-image grey scales for better visibility. (e)~Wider $\vvar{z_1}$ projective fields with normal $\vvar{z_1}$. (f)~ Active $\vvar{z_2}$ dimensions increase with inclreasing $\mathrm{dim}(\vvar{z_2})$ in skipVAE but not in TDVAE. (g)~TDVAE learns complete linear bases in $\vvar{z_1}$, irrespective of $\mathrm{dim}(\vvar{z_2}$). SkipVAE moves as many of these to $\vvar{z_2}$ as possible. (h)~A subset of dimensions of ${\rm E}[\qrec(\cond{\vvar{z_2}}{\vvar{x}})]$ are texture family selective in all studied two-layer models. (i)~Linear decoding accuracies of texture family from ${\rm E}[\qrec(\cond{\vvar{z_2}}{\vvar{x}})]$ in response to texture images, calculated with logistic regression. (j)~Characterization of $\vvar{z_2}$ dimensions. A subset of  $\vvar{z_2}$ units is texture selective in all models. In addition, there is a large set of texture non-selective $\vvar{z_2}$ units in TDVAE and SkipVAE models. These are eliminated by a slightly compressing setting of  $\beta_2$ ($1.25$).}
  \label{fig:representation}
\end{figure}

As the lower layer in all of our generative models was linear (cf.~Fig.~\ref{fig:gen_rec}), we could effectively study what each $\vvar{z_1}$ dimension represents through their projective fields computed with a standard latent traversal procedure. In each model we found that the $\vvar{z_1}$ dimensions clearly clustered into two groups, active and collapsed, based on the mean squared intensity of their projective fields. In image reconstruction experiments, collapsed $\vvar{z_1}$ dimensions were responsible for less than $10^{-6}$ of the generated pixel variances, i.e., their contribution was negligible. For this reason, unless noted otherwise, we are only concerned with active $\vvar{z_1}$ dimensions in the rest of the paper. The number of active $\vvar{z_1}$ dimensions in all LinearVAE, TDVAE, and ChainVAE models was equal to the data dimensions (314 and 1256 in $20 \times 20$ and $40 \times 40$ image patches, respectively), making the active $\vvar{z_1}$ dimensions a complete linear basis in the space of training images. In contrast, the active $\vvar{z_1}$ dimensions in SkipVAE models always formed an undercomplete basis only (Fig.~\ref{fig:representation}c).

We found that the qualitative character of the projective fields of active $\vvar{z_1}$ dimensions depends on the probability distributions chosen for the $\vvar{z_1}$ generative and recognition models. The sparse Laplace distribution leads to localized, oriented, bandpass, Gabor-like filters with low uncertainties (Gabor wavelets are commonly defined by having the lowest possible uncertainty value), reminiscent of the Gabor-like filters found with single-layer sparse linear models of natural images \cite{olshausen1996emergence,barello2018sparse} (see Fig.~\ref{fig:representation}d), while the normal distribution results in more extended oriented filters, akin to the Fourier PCA components of natural images (see Fig.~\ref{fig:representation}e).\footnote{We found one exception to this rule: a minority of $\vvar{z_1}$ dimensions in TDVAE40 and TDVAE20 display nonlocal projective fields with high uncertainties, which are accompanied by a large number of texture-non-selective active dimensions in  $\vvar{z_2}$. With a slight increase of $\beta_2$, as in model TDVAE40125, all $\vvar{z_1}$ filters become G\'abor-like and all active $\vvar{z_2}$ dimensions become texture-selective. For details, see the Supplementary Material.}

\subsection{High-level representation}
\label{sec:higher_level_repr}

We studied the nonlinear $\vvar{z_2}$ representation in our models with the help of the texture dataset discussed in Sec.~\ref{sec_meth}. For each model, the $\vvar{z_2}$ dimensions clustered into an active and a collapsed group, the latter characterized by small variances of posterior $\vvar{z_2}$ means as well as close to unit means of posterior $\vvar{z_2}$ variances within each texture family. From now on, unless denoted otherwise, we are only concerned with active $\vvar{z_2}$ dimensions in the rest of the paper. Training models with different numbers of $\vvar{z_2}$ dimensions revealed that while there is a limit to the number of active $\vvar{z_2}$ dimensions in TDVAE models, SkipVAE models use all $\vvar{z_2}$ dimensions (Fig.~\ref{fig:representation}f). At the same time, while TDVAE and ChainVAE models learn a complete linear $\vvar{z_1}$ representation regardless of the number of available $\vvar{z_2}$ dimensions, in SkipVAE models the sum of active $\vvar{z_1}$ and $\vvar{z_2}$ dimensions is constant and well approximates the data dimensions (Fig.~\ref{fig:representation}f--g). This reveals a fundamental difference between the representations learned by models with Markovian and skip-connection generative models: while the Markovian structure and the linear--nonlinear layering of the Makovian generative models forces all low-level linear features into $\vvar{z_1}$, cleanly separating them from the nonlinear, possibly higher-level features in $\vvar{z_2}$, the generative skip connection in SkipVAE models enables the distribution of a complete low-level representation between $\vvar{z_1}$ and $\vvar{z_2}$ which the SkipVAE models max out in the $\vvar{z_2}$ direction. This reduces both the interpretability of SkipVAE models compared to TDVAE and ChainVAE models and the efficiency of inductive biases to shape the representations.

The synthetic texture images were constructed in a way that texture family cannot be linearly decoded from them (see Fig.~\ref{fig:representation}i, grey bar). It is, however, a robust feature of all inspected hierarchical models that some $\vvar{z_2}$ dimensions linearly encode high-level texture family information (see Fig.~\ref{fig:representation}h--i for examples). The robustness of this property is further supported by the fact that compressing the $\vvar{z_2}$ representation with $\beta_2 > 1$ eliminates the texture-non-selective $\vvar{z_2}$ dimensions but keeps texture-selective ones (see Fig.~\ref{fig:representation}j for examples).

\subsection{Structured correlations at the low-level representation}
\label{sec:corrs}

\begin{figure}
  \centering
  \includegraphics[width=\textwidth]{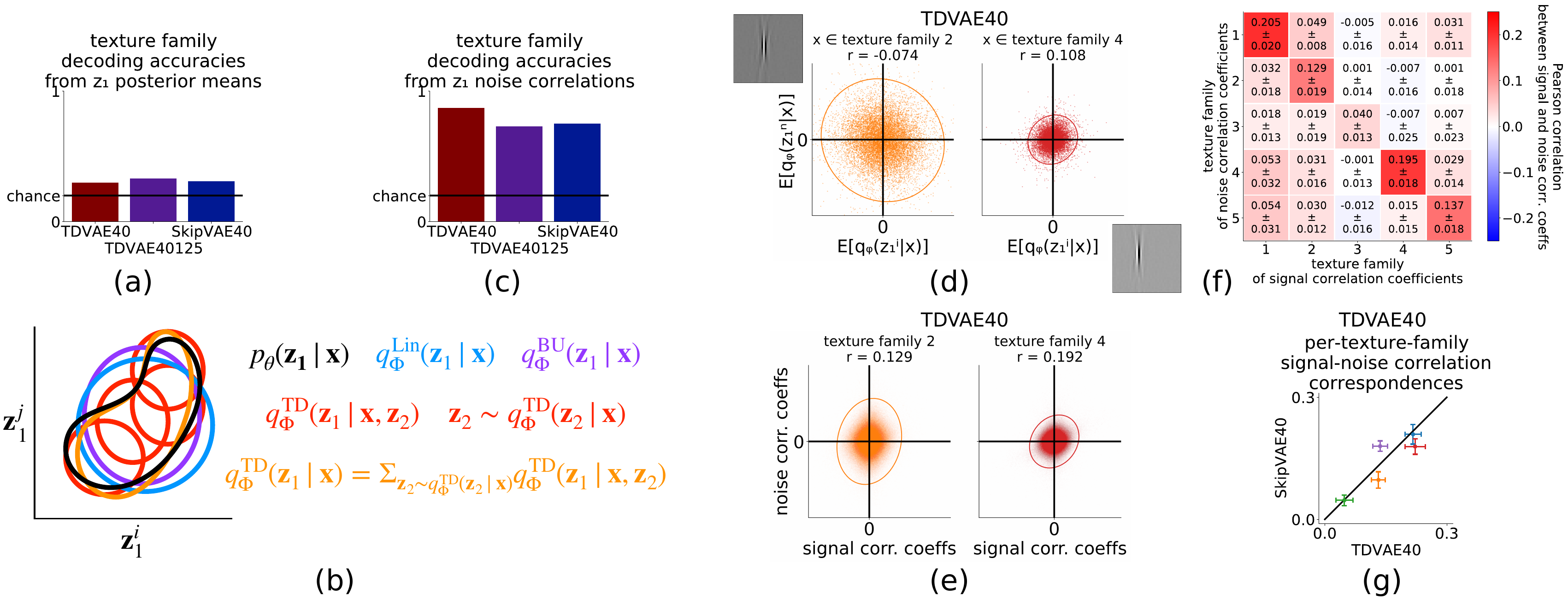}
  \caption{Representation of higher order moments at the level of $\vvar{z_1}$. (a)~Texture family cannot be linearly decoded from ${\rm E} [\qrec(\cond{\vvar{z_1}}{\vvar{x}})]$. (b)~Out of our models, only those with top-down recognition model can represent nontrivial posterior distributions in $\vvar{z_1}$. (c)~Texture family can be linearly decoded from $\vvar{z_1}$ noise correlations for models with a top-down inference structure. (d)~${\rm E}[\qrec(\cond{\vvar{z_1}}{\vvar{x}})]$ for two nearby $\vvar{z_1}$ filters $i$ and $n$ for all stimuli (\emph{dots}) in texture families 2 and 4 in TDVAE40. Signal correlation, $r$, is characteristic to the filter pair and the texture family. (e)~Signal correlation coefficients and stimulus-averaged noise correlation coefficients between each pair of localized $\vvar{z_1}$ filters, within two texture families. Signal and noise correlation coefficients are positively correlated. (f)~Correlations between $\vvar{z_1}$ signal and noise correlation coefficients are significantly stronger within than across texture families. (g)~Signal--noise correlation correspondences do not depend on the presence of generative skip connections. Means and errors are mean and s.d.\ of five repetitions on different balanced subsets of texture images and randomly sampled sets of 100 $\vvar{z_1}$ dimensions with localized projective fields.}
  \label{fig:corrs}
\end{figure}

We have seen above that all hierarchical models studied in this paper represent higher-order statistics of images in $\vvar{z_2}$ which we tested through the linear decodability of texture families from ${\rm E} [\qrec(\cond{\vvar{z_2}}{\vvar{x}})]$ (Figs.~\ref{fig:representation}h--i). But are these high-level image statistics represented in $\vvar{z_1}$ in some way? Since $\vvar{z_1}$ is in a linear generative relation with images $\vvar{x}$, it is not surprising that texture families are not decodable from ${\rm E} [\qrec(\cond{\vvar{z_1}}{\vvar{x}})]$ (Fig.~\ref{fig:corrs}a) where $\qrec(\cond{\vvar{z_1}}{\vvar{x}})$ can be calculated as in Eq.~(\ref{eq:def_noise_corr}). Since $\vvar{z_2}$ is assumed to contribute to shaping the posterior of $\vvar{z_1}$, it is tempting to investigate if higher moments of $\qrec(\cond{\vvar{z_1}}{\vvar{x}})$ carry information about high-level features. By the construction of our models, only those with top-down inference paths (TDVAE, SkipVAE) can represent such higher order moments, those with bottom-up inference (ChainVAE) cannot (Fig.~\ref{fig:corrs}b). Indeed, texture family can be linearly decoded from the posterior correlations of $\vvar{z_1}$ in TDVAE and SkipVAE models (Fig.~\ref{fig:corrs}c):
\begin{eqnarray}
    {\rm corr}^{{\rm noise}}(\vvar{x}) = {\rm corr} [\qrec(\cond{\vvar{z_1}}{\vvar{x}})]  = {\rm corr} \left[\int d\vvar{z}_2 \qrec(\cond{\vvar{z_1}}{\vvar{x}, \vvar{z_2}}) \cdot \qrec(\cond{\vvar{z_2}}{\vvar{x}}) \right],
    \label{eq:def_noise_corr}
\end{eqnarray}
which we call noise correlations following the terminology widespread in the neuroscience literature.

Since model parameters were learned on training data, it is plausible that the uncertainties expressed in $\vvar{z_1}$ posteriors should reflect stimulus properties. \cite{karklin2009emergence} highlighted that texture families possess characteristic correlations in the means of linear filter activations
\begin{eqnarray}
    {\rm corr}^{{\rm signal}}(\rm tf) = {\rm corr}_{p_{\rm tf}(\vvar x)} [{\rm E}[\qrec(\cond{\vvar{z_1}}{\vvar{x}})]],
    \label{eq:def_signal_corr}
\end{eqnarray}
termed signal correlations ($\rm tf$ denotes texture family). This is also characteristic to all of our models, too (Fig.~\ref{fig:corrs}d). We found that, within each texture family, the elements of the signal correlation and the texture-averaged noise correlation matrices (corresponding to filter pairs) are positively correlated (Fig.~\ref{fig:corrs}e) and that this correlation is significantly stronger within than across texture families (Fig.~\ref{fig:corrs}f). Within-family signal--noise correlation correspondences do not seem to depend on the presence of skip connections in the generative model (Fig.~\ref{fig:corrs}g). Note, that noise correlations cannot be represented in the ChainVAE model architecture, therefore the correspondence between signal and noise correlations is not achievable for that model.

\subsection{Image inpainting experiments}
\label{sec:masking}

We demonstrated in earlier sections that our models with a top-down inference path 1) discover and represent higher-order statistical properties of images $\vvar{x}$ in $\vvar{z_2}$ through their recognition model $\qrec(\cond{\vvar{z_2}}{\vvar{x}})$, and 2) convey this information to $\vvar{z_1}$ through their recognition model $\qrec(\cond{\vvar{z_1}}{\vvar{x}, \vvar{z_2}})$ in the form of higher-order-statistics dependent correlations. Our next question is whether our models utilize this information when reconstructing images through $p_{\theta}({\bf x}\,|\, {\bf z}_1)$ or $p_{\theta}({\bf x}\,|\, {\bf z}_1, {\bf z}_2)$. We investigate this question with an image inpainting experiment with masked images.

Since our models are not trained specifically for image inpainting tasks, we should not compare them to e.g.\ state of the art deep generative models for image completion \cite{harvey2021image}. But if TDVAE and SkipVAE models would surpass LinearVAE models in image inpainting tasks, it would indicate that they can extract high-level contextual information even from corrupted (masked) images and introduce this to the corrupted (masked) image area in a meaningful way.

\begin{figure}
  \centering
  \includegraphics[width=\textwidth]{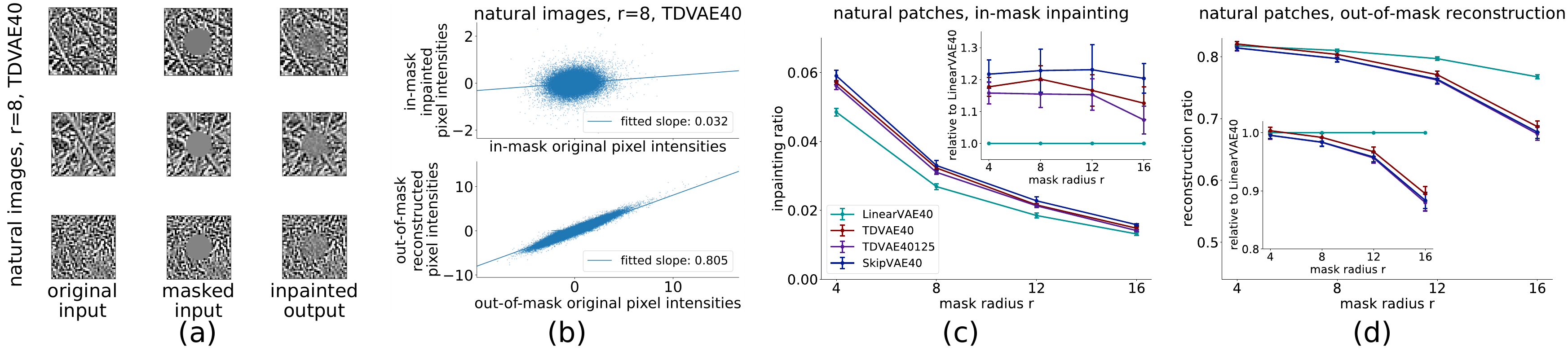}
  \caption{(a)~Setup of masking experiment. (b)~Inpainting and reconstruction slopes were calculated with linear fits to in-mask inpainted and out-of-mask reconstructed vs original pixel intensities ({\em inpainting and reconstruction ratios}). (c)~Top-down inference substantially enhances in-mask image completion. (d)~Top-down inference somewhat degrades out-of-mask reconstruction.}
  \label{fig:masking}
\end{figure}

We started our image inpainting experiments with masking out disks of various radii $r$ in the centers of natural images, inferring the latent representation $\vvar{z_1}$ and $\vvar{z_2}$ of these masked images with our models, and generating a (hopefully) inpainted image from $\vvar{z_1}$ and $\vvar{z_2}$ (Figs.~\ref{fig:masking}a). We characterized in-mask inpainting and out-of-mask reconstruction performance with the slopes of lines fitted to inpainted/reconstructed vs original pixel intensities ({\em inpainting/reconstruction ratio}, Figs.~\ref{fig:masking}b). To test robustness, we repeated this procedure on five non-overlapping natural image subsets for each mask radius $r$ and report the means and standard deviations of these repetitions in Figs.~\ref{fig:masking}c--d.

Fig.~\ref{fig:masking}c demonstrates that top-down inference substantially enhances in-mask image inpainting, especially at small mask radii $r$ where a large non-masked image area can be used to infer higher-order image features. As a control Fig.~\ref{fig:masking}d shows that top-down inference performs somewhat worse than linear reconstruction for out-of-mask image reconstruction. It means that the top-down models fill in the mask in a smoother, more semantic way.

\section{Conclusions}
In this paper we explored a particular flavor of deep generative models, VAEs to learn a hierarchical representation of natural images. Hierarchical forms of VAE are inspired both by neuroscience (unsupervised learning of the cascade of processing stages in the visual cortex) and by machine learning (e.g. representation learning). We found that simple inductive biases, such as linearity of the generative model and Laplacian prior/posterior, contribute to learning a hierarchy of features that are reminiscent to that found in the visual cortex of primates. In particular, a representation for texture-like features emerged that was very robust against variations in the architecture both of the recognition and generative models. However, a variant of the generative model that features skip connections, and has recently gained popularity, was found to show consistent interference between the low-level and high-level representations: dimensionality of the high-level layer constrained the dimensionality of the representation at the low-level layer. This finding makes sense since the dimensionality of the input limits the linear information content and skipVAE has the potential to take over control over the linear components. It is a warning sign too, that the complexity of the generative model prevents the meaningful separation of features.

Our results show that in hVAEs the higher order moments of the posterior are also characteristic of the posterior besides the mean. Importantly, this aspect of the representation distinguishes recognition models that seem to be equivalent: the posterior of the low-level latent layer in chainVAE is incapable of learning higher order moments and is therefore learning a poorer representation. Remarkably, this property of the representation is one of the normative arguments to build hierarchical versions of VAEs. The top-down architecture of the recognition model featured in both TDVAE and skipVAE reflect a fundamental feature of the anatomy of the visual cortex, the rich projections from cortical areas higher up in the hierarchy to the lower areas \cite{gilbert2013top}, establishing a normative argument for their existence and a perspective to investigate their properties. Finally, our masking experiments showed that high-level latent representations can contribute to image inpainting.

We found strong links between the so-called signal correlations and noise correlations in posteriors. Such a link between prior correlations, signal correlations, and noise correlations has been investigated in neuroscience \cite{orban2016neural}, thus these results provide a theoretical framework for predicting patterns in higher-order moments of the neural response statistics.  Interestingly, top-down components of inference have been implicated in shaping noise correlations in the visual cortex of monkeys \cite{banyai2019stimulus}. Our results can establish specific links between receptive field properties, noise correlations, and signal correlations, leading to a better understanding of the neural code.

The chain structured recognition model of the chainVAE model is a formulation that is considered as an alternative to the top-dwon recognition model \cite{havtorn2021hierarchical} and has been used as a building block for VAEs \cite{burda2015importance}. The learned representations at the level of posterior correlations has not been investigated before and therefore the lack of expressive power has not been identified before. This limitation can be overcome by learning the covariance structure of the variational posterior. It is the subject of future research to see if the ELBO can efficiently support learning a similar covariance structure as the one obtained through the top-down version of the recognition model.

The size of the image patch we used limits the complexity of the features learned and therefore the added benefit of deeper generative models. In future work we seek to explore opportunities in increased patch sizes. We used fully connected MLPs for a better transparency of the model but using convolutional layers instead will provide a margin for training the model with larger image patches, which will also give an opportunity to explore the response properties in higher visual areas.

\begin{ack}
Funding in direct support of this work: Human Frontier Science Program (RGP0044/2018), and the  Artificial Intelligence National Laboratory of Hungary (NKFIH-1530-4/2021).

Additional revenues related to this work: none.
\end{ack}

\section*{}
\medskip

{
\small

\bibliographystyle{plain}
\bibliography{arxiv_org}
}

\newpage

\appendix

\section{Appendix}

\subsection*{Texture data}

\begin{table}
  \caption{Texture family seed images and preprocessing parameters. For details, see text.}
  \label{tab:texture_stimuli}
  \centering
  \begin{tabular}{llllr}
    \toprule
    Fig.~\ref{fig:texture_stimuli} panel & Origin & Filename & Channel & Subsampling \\
    \midrule
        (a) & \href{https://www.textures.com/}{textures.com} & FoodGrains0001\_1\_seamless\_S & green & $4 \times 4$ \\
        (b) & \href{https://www.textures.com/}{textures.com} & Leather0028\_1\_M & green & $2 \times 2$ \\
        (c) & \href{https://www.textures.com/}{textures.com} & SoilCracked0079\_1\_seamless\_S & red & $2 \times 2$ \\
        (d) & \href{https://www.textures.com/}{textures.com} & Carpet0025\_1\_seamless\_S & blue & $2 \times 2$ \\
        (e) & \cite{brodatz-1966-textures} & D111 & N/A & N/A \\
    \bottomrule
  \end{tabular}
\end{table}

\begin{figure}
  \centering
  \includegraphics[width=\textwidth]{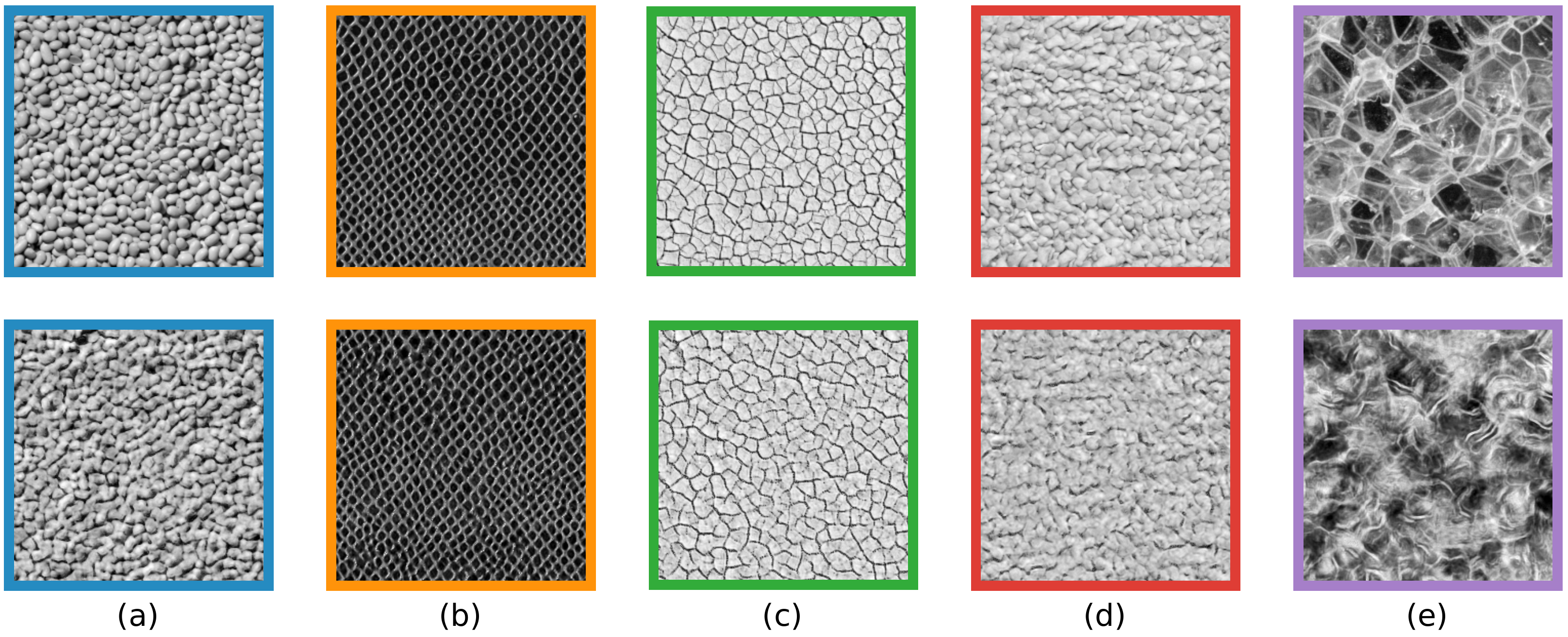}
  \caption{Top row: $256 \times 256$ crops from the five texture family seed images after preprocessing with the parameters in Table~\ref{tab:texture_stimuli}. Bottom row: $256 \times 256$ crops from textures synthesized with \cite{portilla2000parametric}.}
  \label{fig:texture_stimuli}
\end{figure}

Table~\ref{tab:texture_stimuli} shows the origin and the preprocessing parameters for the texture family seed images used for synthesizing the texture families that we used in model testing. The colored seed images (a)--(d) were turned into greyscale by selecting one of their color channels, then cubic subsampling was applied with a window size that ensured that the dominant Fourier component fits into a $20 \times 20$ cropped image patch. The greyscale seed image (e) has no characteristic Fourier component (it has a scale-free autocorrelation function), that is why no subsampling was applied. The seed images preprocessed in this way are shown in the top row of Fig.~\ref{fig:texture_stimuli}. (For information on seed image licenses, see Sec.~\ref{sec_meth}.)

The preprocessed texture seed images were then fed into the texture synthesis method \cite{portilla2000parametric} one by one to generate a large number of texture images for each texture family (for samples, see the bottom row of Fig.~\ref{fig:texture_stimuli}). These synthesized  texture images were then used to generate a large number of $20 \times 20$ and $40 \times 40$ pixel whitened texture patches, as described in Sec.~\ref{sec_meth}. To promote statistical correspondence to our natural training images, we only used texture families on which a sparse LinearVAE model (see Tab.~\ref{tab:models}) learned a complete basis of localized, oriented, bandpass filters.

\begin{table}
  \caption{Parameters and download URLs for the training and test datasets used in the paper. See text for details.}
  \label{tab:datasets}
  \centering
  \begin{tabular}{lrl}
    \toprule
    Image type & Patch size & Link \\
    \midrule
        natural & $20 \times 20$ & \href{https://zenodo.org/record/6576020/files/fakelabeled_natural_commonfiltered_640000_20px.pkl?download=1}{download} \\
        natural & $40 \times 40$ & \href{https://zenodo.org/record/6576020/files/fakelabeled_natural_commonfiltered_640000_40px.pkl?download=1}{download} \\
        textures & $20 \times 20$ & \href{https://zenodo.org/record/6576020/files/labeled_texture_oatleathersoilcarpetbubbles_commonfiltered_640000_20px.pkl?download=1}{download} \\
        textures & $40 \times 40$ & \href{https://zenodo.org/record/6576020/files/labeled_texture_oatleathersoilcarpetbubbles_commonfiltered_640000_40px.pkl?download=1}{download} \\
    \bottomrule
  \end{tabular}
\end{table}

The ensuing training and test datasets were placed into a public repository that ensures long-term preservation of the data, provides a Digital Object Identifier, and publishes metadata in several metadata standards, including Schema.org and DCAT. In the interest of anonymity, though, we do not give access to this repository yet, and provide only anonymyzed links in Table~\ref{tab:datasets}. Each file is in the pickle format and was generated with Python 3.8.5. Each file contains a Python dictionary with the following fields:
\begin{description}
\item[`train\_images'] 640,000 float32 images used for model training. $20 \times 20$ pixel images contain 400 pixel intensities, $40 \times 40$ pixel images contain 1600 pixel intensities each.
\item[`train\_labels'] float32 labels for each image in \textbf{`train\_images'}. All natural images are labeled with 0.0. Texture images are labeled with 0.0, 1,0, 2.0, 3.0, or 4.0, according to their texture family.
\item[`test\_images'] 64,000 float32 images used for model testing. $20 \times 20$ pixel images contain 400 pixel intensities, $40 \times 40$ pixel images contain 1600 pixel intensities each.
\item[`test\_labels'] float32 labels for each image in \textbf{`test\_images'}. All natural images are labeled with 0.0. Texture images are labeled with 0.0, 1,0, 2.0, 3.0, or 4.0, according to their texture family.
\end{description}
These data are published under the terms of the \href{https://creativecommons.org/licenses/by/4.0/legalcode}{Creative Commons Attribution 4.0 International} license.

\subsection*{Details of model training}

\subsubsection*{Architectural details}
We first discuss the details of the architectural choices for the top-down recognition model present in TDVAE and SkipVAE. There are four MLPs defined for the recognition models as it is depicted in Fig. \ref{fig:gen_rec}(e). The first neural network (\emph{MLP.a}) maps the pixel space to a layer $L_x$ which is shared between the computation of $\qrec(\cond{\vvar{z_2}}{\vvar{x}})$ and $\qrec(\cond{\vvar{z_1}}{\vvar{x}, \vvar{z_2}})$. From $L_x$ \emph{MLP.b} computes the mean and variances of the $\qrec(\cond{\vvar{z_2}}{\vvar{x}})$ distribution. The third MLP (\emph{MLP.c}) transforms $\vvar{z_2}$ to the layer $L_z$. We fuse the information from $\vvar{x}$ and $\vvar{z_2}$ by concatenating $L_x$ and $L_z$ and apply an MLP on the combined layer to get the mean and variances of $\qrec(\cond{\vvar{z_1}}{\vvar{x}, \vvar{z_2}})$ (\emph{MLP.d}).
The number of hidden layers and hidden units used in each MLP for calculating the means and the standard deviations of the conditional generative and variational posterior distributions is shown in Table~\ref{tab:NNparamschain} for models with bottom-up recognition models and in Table~\ref{tab:NNparamsTD} for models with top-down recognition model.

\begin{table}
  \caption{Number of hidden units in each MLP layer computing the mean and the standard deviation of each conditional distribution in the bottom-up recognition models.}
  \label{tab:NNparamschain}
  \centering
  \begin{tabular}{llrrr}
    \toprule
    Name & Architecture & $\pgen(\cond{\vvar{z_1}}{\vvar{z_2}})$ &  $\qrec(\cond{\vvar{z_1}}{\vvar{x}})$ & $\qrec(\cond{\vvar{z_2}}{\vvar{z_1}})$ \\
    \midrule
    LinearVAE40   & LinearVAE & N/A & (2000, 2000) & N/A \\
    ChainVAE20    & ChainVAE  & (500) & (500, 500) & (500, 300, 100) \\
    ChainVAE20125    & ChainVAE  & (500) & (500, 500) & (500, 300, 100) \\
    \bottomrule
  \end{tabular}
\end{table}

\begin{table}
  \caption{Number of hidden units in each MLP layer computing the mean and the standard deviation of each conditional distribution in the top-down recognition models.}
  \label{tab:NNparamsTD}
  \centering
  {\footnotesize
  \begin{tabular}{lrrrrrr}
    \toprule
    Name & $\pgen(\cond{\vvar{z_1}}{\vvar{z_2}})$ & MLP.a & MLP.b & MLP.c & MLP.d & skip connection\\
    \midrule
    TDVAE40            & (2000) & (2000) & (1000, 500, 250) & (250, 500, 1000, 2000) & (2000) & N/A \\
    TDVAE40n          & (2000) & (2000) & (1000, 500, 250) & (250, 500, 1000, 2000) & (2000) & N/A \\
    TDVAE40125       & (2000) & (2000) & (1000, 500, 250) & (250, 500, 1000, 2000) & (2000) & N/A \\
    SkipVAE40       & (2000) & (2000) & (1000, 500, 250) & (250, 500, 1000, 2000) & (2000) & (2000, 1800)\\
    TDVAE20           & (500) & (500) & (300, 100) & (100, 300, 500) & (500) & N/A \\
    SkipVAE20       & (500) & (500) & (300, 100) & (100, 300, 500) & (500) & (500, 450)\\
    \bottomrule
  \end{tabular}
  }
\end{table}

\subsubsection*{Model training details}
We trained all of our models with the Adam optimizer \cite{kingma2014adam}. We found that while the learned $\vvar{z_1}$ representation was robust against regularization techniques (we tested weight decay, gradient clipping, and gradient skipping), the learned $\vvar{z_2}$ representation was sensitive to these. To eliminate such regularization artifacts, we turned off weight decay and increased the gradient clipping and skipping thresholds to have an activation frequency below $10^{-6}$. As a final step, we kept decreasing the learning rate (constant in each experiment) until the learned representation in $\vvar{z_2}$ converged. Test ELBOs were at most 20\% higher than training ELBOs, demonstrating that the number of training examples (\num{3.2e5} to \num{6.4e5}) was enough to avoid overfitting. We stopped training at full convergence. The main hyperparameters of a selected subset of experiments is shown in~Table~\ref{tab:models}. Training this set of models took 787~hours altogether on an internal computer sporting an Nvidia GeForce RTX 3080 Ti GPU. The source code of our models contains code from \cite{rao-2019-continual} which uses the Apache-2.0 license.

\begin{table}
  \caption{Model optimization parameters.}
  \label{tab:model_opt}
  \centering
  \begin{tabular}{llr}
    \toprule
    Name & Architecture & Learning rate(s) \\
    \midrule
    LinearVAE40   & LinearVAE & \num{3e-5}\\
    TDVAE40       & TDVAE     & \num{5e-5} $\to$ \num{2.5e-5} \\
    TDVAE40n      & TDVAE     & \num{5e-5}\\
    TDVAE40125    & TDVAE     & \num{5e-5}  \\
    SkipVAE40     & SkipVAE   & \num{2e-5}  \\
    TDVAE20       & TDVAE     & between \num{2.5e-5} and $10^{-4}$  \\
    SkipVAE20     & SkipVAE   & between \num{5e-5} and $10^{-4}$  \\
    ChainVAE20    & ChainVAE  & $10^{-5}$      \\
    ChainVAE20125    & ChainVAE  & $10^{-5}$      \\
    \bottomrule
  \end{tabular}
\end{table}

\end{document}